\documentclass{desyproc}

\begin{document}
\title{APEX: A Prime EXperiment at Jefferson Lab}

\author{{\slshape James Beacham}\\[1ex]
New York University (NYU), New York, NY
}

\contribID{familyname\_firstname}

\desyproc{DESY-PROC-2012-04}
\acronym{Patras 2012} 
\doi  

\maketitle

\begin{abstract}
APEX is an experiment at Thomas Jefferson National Accelerator Facility (JLab) in Virginia, USA, that searches for a new gauge boson ($A^\prime$) with sub-GeV mass and coupling to ordinary matter of $g^\prime \sim (10^{-6} - 10^{-2}) e$.  Electrons impinge upon a fixed target of high-Z material.  An $A^\prime$ is produced via a process analogous to photon bremsstrahlung, decaying to an $e^+ e^-$ pair.  A test run was held in July of 2010, covering $m_{A^\prime}$ = 175 to 250 MeV and couplings $g^\prime/e \; \textgreater \; 10^{-3}$.  A full run is approved and will cover $m_{A^\prime} \sim$ 65 to 525 MeV and $g^\prime/e \; \textgreater \; 2.3 \times10^{-4}$.
\end{abstract}

\section{Motivations}

The Standard Model (SM) of particle interactions is described by an SU(3)$_{C}\times$SU(2)$_{L}\times$U(1)$_{Y}$ gauge group, where the forces are mediated by vector bosons.  For an extension of this model to have thus far evaded detection the corresponding gauge boson must have a mass of $\mathcal{O}$(TeV) or must be very weakly coupled to ordinary matter, with a coupling strength $g^\prime$ suppressed relative to the electromagnetic charge $e$ by $\epsilon \equiv g^\prime/e \sim 10^{-6} - 10^{-2}$~\cite{Essig:2009nc} (or, equivalently, $\alpha^\prime / \alpha = \epsilon^{2}$).  This new gauge boson, $A^\prime$, corresponding to a U(1)$^\prime$ extension of the SM can acquire an effective interaction with electromagnetism via kinetic mixing, where quantum loops of arbitrarily heavy particles provide a means by which the hidden U(1)$^\prime$ sector couples to the visible sector; see, e.g.,~\cite{Holdom:1985ag,Galison:1983pa,Strassler:2006im}.  The possibility of the existence of an $A^\prime$ with a small EM charge can be tested at fixed target facilities such as the Thomas Jefferson National Accelerator Facility (JLab).  APEX, The A Prime EXperiment, searches for an $A^\prime$ at JLab and is described in brief here. For a full description of the experiment see~\cite{Essig:2010xa} and for a more detailed description of the results of the test run see~\cite{Abrahamyan:2011gv}.

In addition to the general interest in discovering an extension of the SM, a hidden gauge sector with a gauge boson with mass in the MeV to GeV range could address dark matter anomalies and the anomalous magnetic moment of the muon.  For a complete discussion of these possibilities, see~\cite{Bjorken:2009mm}.

\section{Existing Constraints}

\begin{wrapfigure}[22]{L}{0cm}
\centering
\includegraphics[width=0.6\textwidth]{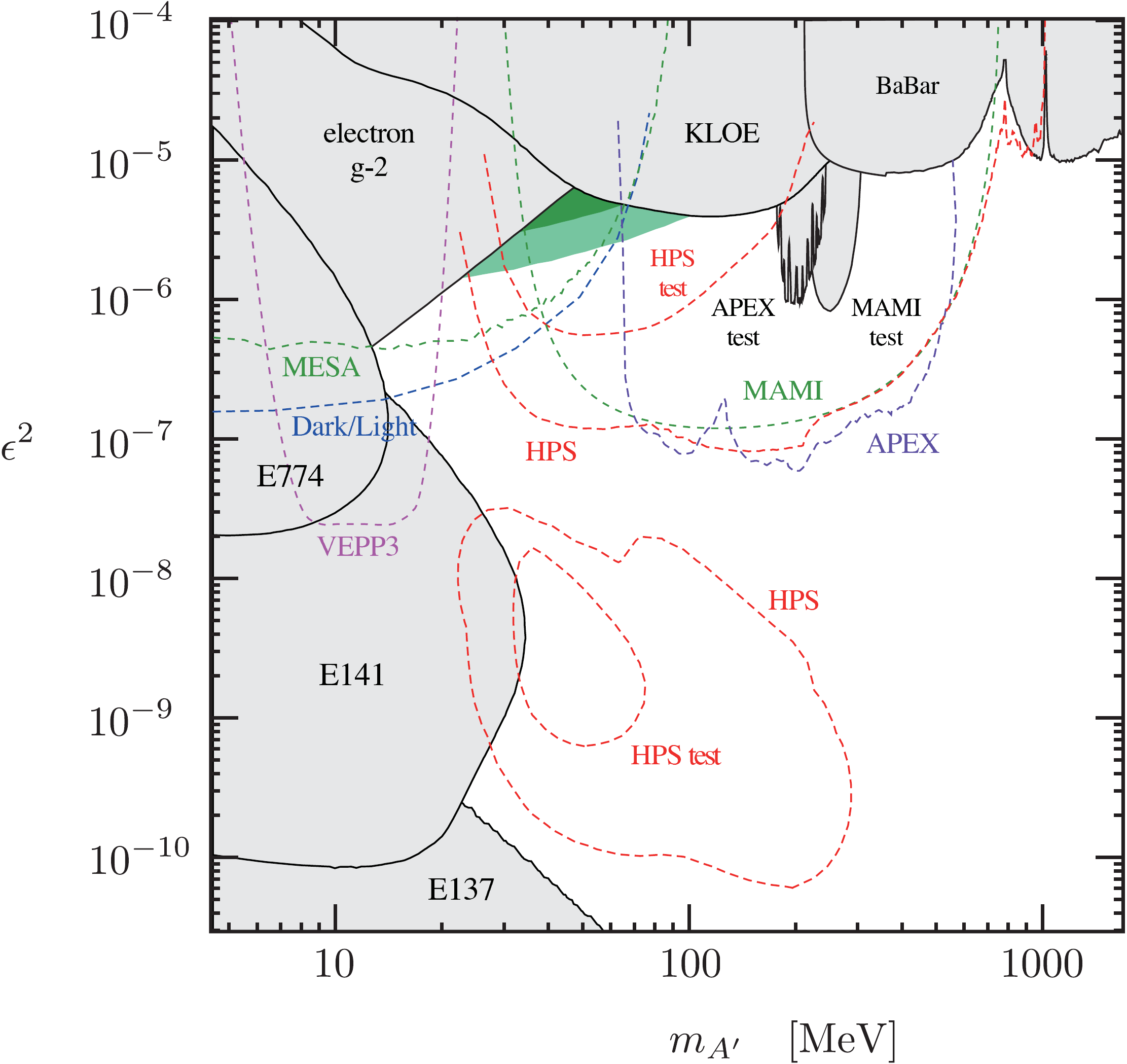}
\caption{Existing and planned constraints in the $\epsilon - m_{A^\prime}$ plane, as of late 2012. From~\cite{Endo:2012hp}.}\label{Fig:epsmA}
\end{wrapfigure}

Aside from these suggestive motivations, the coupling strength and mass of the $A^\prime$ are not predicted.  Thus, searches for this new gauge boson must be conducted over wide ranges of both.  As a result, prior to 2009, the areas of parameter space probed by APEX were remarkably weakly constrained.  Following the observation~\cite{Bjorken:2009mm} that much of this range could be probed at existing experimental facilities, a renewed interest in such experiments has led to the current constraints and planned experimental sensitivities shown in Figure~\ref{Fig:epsmA}.  For a complete description of these constraints, see~\cite{Endo:2012hp} and references therein.

As seen in Figure~\ref{Fig:epsmA}, APEX covers a large portion of this area of parameter space, from $m_{A^\prime} \sim$ 65 to 525 MeV and with coupling reach to $g^\prime/e \; \textgreater \; 2.3 \times 10^{-4}$.  A test run for APEX was performed in July of 2010 and demonstrated the feasibility of the full experiment.

\section{APEX at Jefferson Lab's Hall A}

APEX is designed to take full advantage of JLab's Continuous Electron Beam Accelerator Facility and the two High Resolution Spectrometers (HRSs) in Hall A.  For the test run, an electron beam of energy 2.260 GeV and an intensity of up to 150 $\mu$A was used, incident upon a tantalum foil of thickness 22 mg/cm$^{2}$.  The central momentum of each HRS was $\simeq$ 1.131 GeV with a momentum acceptance of $\pm$ 4.5\%.

\begin{wrapfigure}[11]{L}{0cm}
\centering
\subfigure[]{
\includegraphics [trim = 0mm 5mm 0mm 0mm, width = 0.19\textwidth]{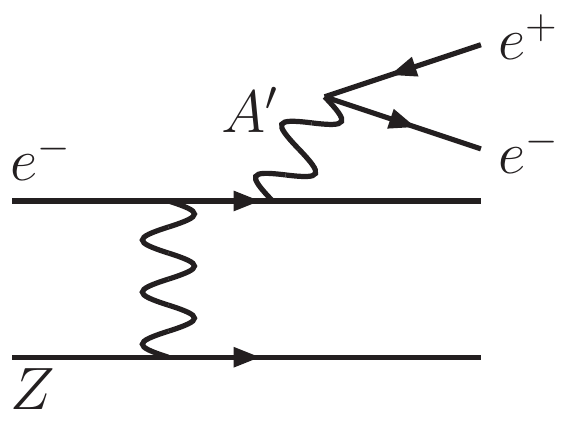}
\label{fig:production.1a}
}
\subfigure[]{
\includegraphics [trim = 0mm 5mm 0mm 0mm, width = 0.19\textwidth]{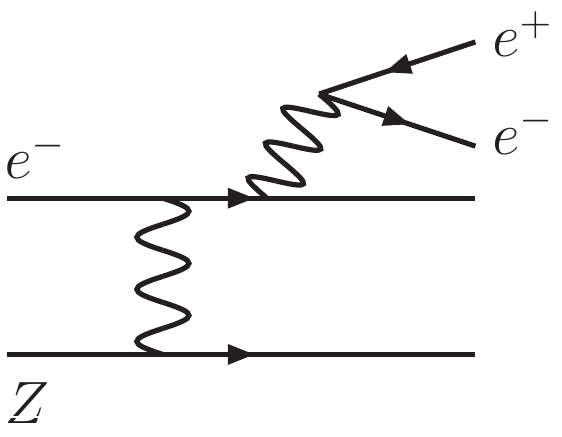}
\label{fig:production.1b}
}
\subfigure[]{
\includegraphics [trim = 0mm 5mm 0mm 0mm, width = 0.20\textwidth]{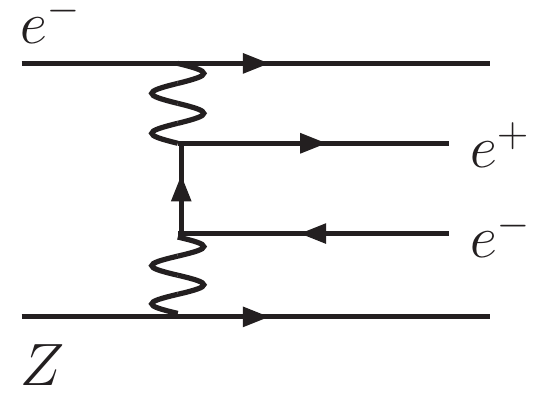}
\label{fig:production.1c}
}
\vspace{-.10 in}
\caption{$A^\prime$ signal process (a) and irreducible QED backgrounds (b) and (c).
}
\label{Fig:diagrams}
\vspace{-.30 in}
\end{wrapfigure}

An $A^\prime$ is produced via a process analogous to photon bremsstrahlung and decays to an $e^+ e^-$ pair; thus, the $A^\prime$ signal will appear as a small, narrow bump in the invariant mass spectrum of $e^+ e^-$ pairs from background QED processes.  The diagrams for signal and irreducible backgrounds are shown in Figure~\ref{Fig:diagrams}.

The opening angle $\Theta_{0}$ of the $e^+ e^-$ pair is set by $m_{A^\prime}$ and the incident electron beam energy as $\Theta_{0} \sim m_{A^\prime}/E_{b} \approx 5^{\circ}$, with no such expectation for the QED backgrounds.  This motivates a symmetric HRS configuration with both spectrometer arms positioned far forward.  To optimize sensitivity to $A^\prime$ decays, dipole septum magnets are placed between the target and the HRS aperture.

The Hall A HRSs consist of several different components to allow for the measurement of the position and momentum of charged particles to a high degree of accuracy.  Vertical drift chambers allow for an accurate determination of the full 3D track of an incoming particle.  Two separate sets of scintillators provide timing information, to identify coincident $e^+ e^-$ pairs.  Online particle ID is provided by a gas Cherenkov detector and a lead glass calorimeter allows for further offline rejection of pion backgrounds.  A sieve-slit method is used for optics calibration.

Excellent mass resolution is required to enable the identification of an $A^\prime$ resonance.  The HRSs are designed to achieve high momentum resolution at the level of $\delta p / p \sim 10^{-4}$, providing a negligible affect upon the mass resolution.  Angular resolution and multiple scattering in the target are the dominant contributions to the mass resolution, as shown in Table~\ref{tab:massres}.

\begin{wraptable}{r}{0.55\textwidth}
\centering
\begin{tabular}{| c | c | c | c |}
\hline
mrad  & Optics & Tracking & MS in target \\
\hline
$\sigma$(horiz)  & 0.11 & $\sim$0.4 & 0.37 \\
\hline
$\sigma$(vert)  & 0.22 & $\sim$1.8 & 0.37 \\
\hline
\end{tabular}
\caption{Contributions to APEX mass resolution.}
\label{tab:massres}
\end{wraptable}

\noindent For the test run, APEX achieved a mass resolution of $\sigma \sim 0.85 - 1.11$ MeV, varying over the full $m_{A^\prime}$ range.

Reducible backgrounds, including electron or proton singles, pions, accidental $e^+ e^-$ coincidences, and $e^+ e^-$ pairs from real photon conversions, are rejected using a combination of different triggers.

The final event sample trigger for the test run required a double coincidence gas Cherenkov signal within a 12.5 ns window in each arm.  The resulting data sample consisted of 770,500 true $e^+ e^-$ coincident events with 0.9\% (7.4\%) meson (accidental $e^+ e^-$ coincidence) contamination.  The final data sample forms the basis of an invariant mass spectrum of $e^+ e^-$ pairs.  A bump hunt for a small, narrow resonance was performed, using the profile likelihood ratio as the test statistic.

\section{Test Run Results}

\begin{figure}
\centering
\begin{minipage}{0.5\textwidth}
\centering
\includegraphics[width=0.8\linewidth]{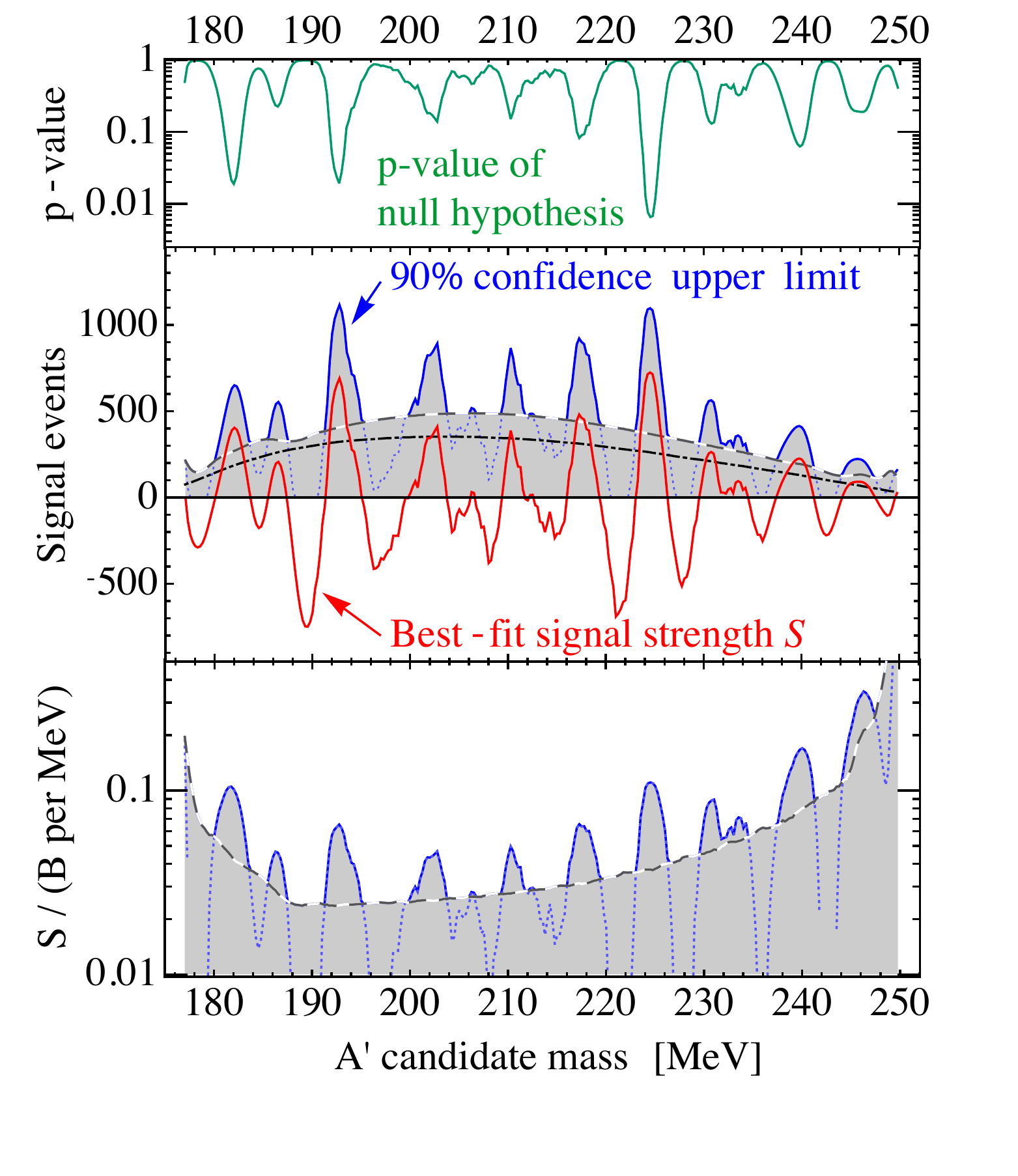}
\caption{Results of the resonance search.}
\label{Fig:results}
\end{minipage}%
\begin{minipage}{0.5\textwidth}
\centering
\includegraphics[width=0.8\linewidth]{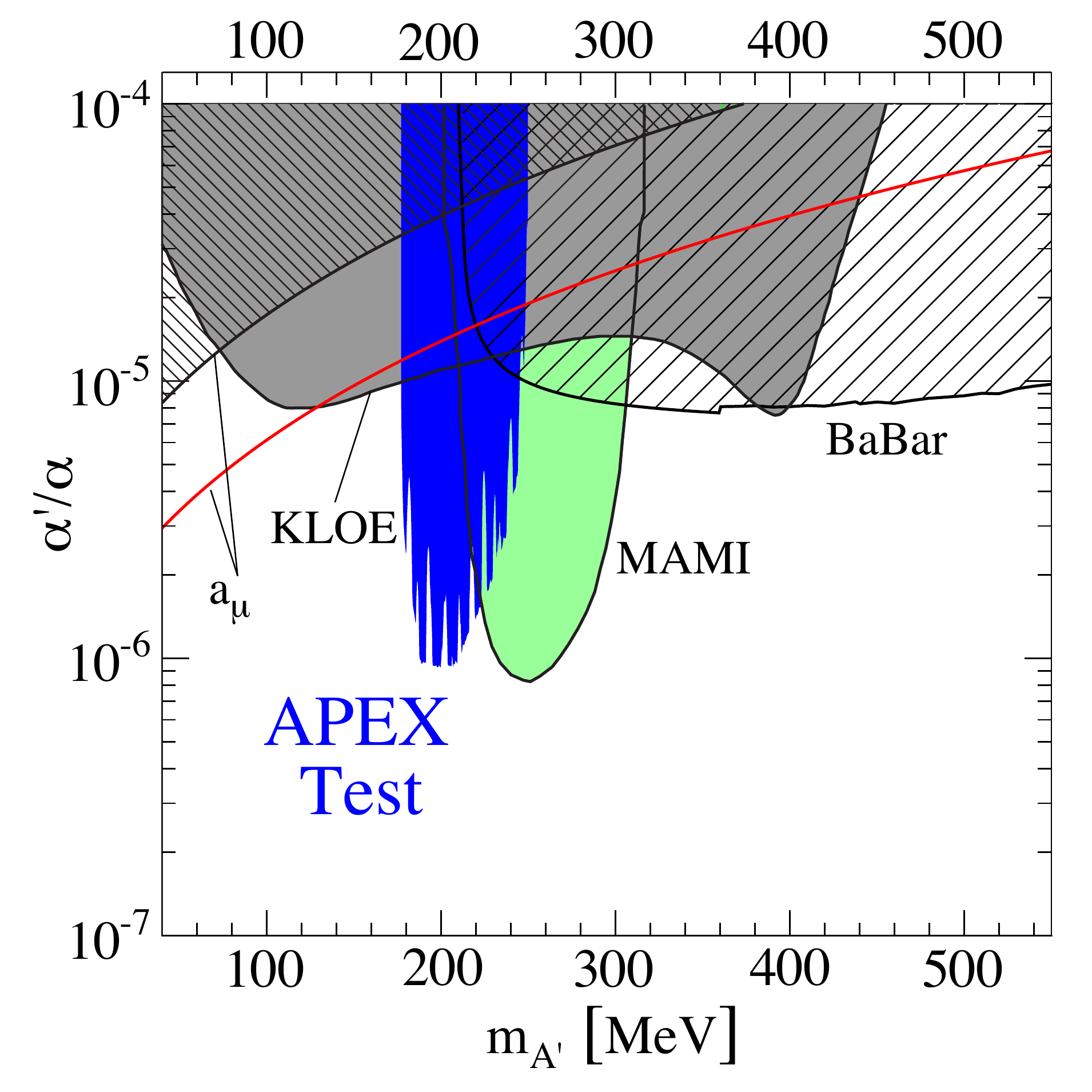}
\caption{Upper limit on coupling.}
\label{Fig:coupling}
\end{minipage}
\end{figure}

No significant excess was found over the invariant mass range of $m_{A^\prime}$ = 175 to 250 MeV; see Figure~\ref{Fig:results}.  The most significant excess was at 224.5 MeV with a p-value of 0.06\%.  Out of $\sim$1000 pseudoexperiments based on the test run data, 40\% yielded a p-value at least as extreme as 0.06\% somewhere in the mass range.

The upper limit on number of signal events, $S$, compatible with a background fluctuation at the 90\% CL was translated into an upper limit on the $A^\prime$ coupling, $\alpha^\prime / \alpha$, by exploiting the kinematic similarities between $A^\prime$ and $\gamma^{*}$ production~\cite{Bjorken:2009mm}.  Based on Monte Carlo simulations, the ratio $f$ of the radiative-only cross section to the full QED background cross section varies linearly from 0.21 to 0.25 across the APEX mass range and, thus, all backgrounds can be normalized to the radiative background.  The final expression relating $S_{max}$ and $(\alpha^\prime / \alpha)_{max}$ is

\begin{equation}
\left(\frac{\alpha'}{\alpha}\right)_{max} = \left(\frac{S_{max}/m_{A'}}{f \cdot \Delta B / \Delta m}\right) \times \left(\frac{2 N_{\text{eff}} \alpha}{3 \pi}\right),
\nonumber
\end{equation}

\noindent and the upper limit on coupling is shown in Figure~\ref{Fig:coupling}.

\section{Full Run Plans}

The APEX full run is approved and will be ready to run when JLab switches on in 2014 after an upgrade of the beam energy from 6 to 12 GeV.  The full run will take data for $\sim$34 days at four different energy and spectrometer settings, and will cover a larger mass range, $m_{A^\prime}$ = 65 to 525 MeV, using a 50 cm long multifoil target.  The full run statistics will be $\sim$200 times larger than the test run, allowing sensitivity to $\alpha^\prime / \alpha$ 1-2 orders of magnitude below current limits.   A new optics calibration method is currently being tested and data acquisition rates are being improved, to allow for up to 5 kHz.   A complete description of the full run is in~\cite{Essig:2010xa}.

\section{Acknowledgments}

The author would like to acknowledge Rouven Essig, Philip Schuster, Natalia Toro, and Bogdan Wojtsekhowski, the APEX spokespeople, and Sergey Abrahamyan, Eric Jensen, Jin Huang, and Kyle Cranmer, the rest of the members of the core analysis team for the APEX test run, for their fruitful collaboration.  Additional thanks are due the Hall A collaboration and the Jefferson Lab staff for their outstanding support.


\begin{footnotesize}

\end{footnotesize}


\end{document}